\documentclass[aps,prl,showpacs]{revtex4}
\usepackage{graphicx}
\usepackage{hyperref}
\usepackage{amsfonts}
\usepackage{amsbsy}
\usepackage{amssymb}
\usepackage{amsmath}
\usepackage{xcolor}
\usepackage[normalem]{ulem}
\usepackage{float}
\usepackage{multirow}
\usepackage{amssymb}  
\usepackage{amsthm}
\usepackage{amsmath}

\begin{document}

\title{Nonlinear dynamical systems: Time reversibility {\it versus} sensitivity to the initial conditions} 
\author{Constantino Tsallis$^{1,2,3}$}
\email{tsallis@cbpf.br}
\author{Ernesto P. Borges$^{4}$}
\email{ ernesto@ufba.br}
\affiliation{$^1$Centro Brasileiro de Pesquisas Fisicas \\
and National Institute of Science and Technology for Complex Systems\\
\mbox{Rua Xavier Sigaud 150, Rio de Janeiro 22290-180, Brazil} \\
 $^2$ Santa Fe Institute, 1399 Hyde Park Road, Santa Fe, 
 New Mexico 87501, USA \\
 $^3$ Complexity Science Hub Vienna, Josefst\"adter Strasse 
 39, 1080 Vienna, Austria \\
 $^4$Instituto de Fisica, Universidade Federal da Bahia\\ 
 and National Institute of Science and Technology for Complex Systems\\
 Salvador-BA 40170-115, Brazil}

\date{\today}

\begin{abstract}
Time reversal of vast classes of phenomena has direct implications with predictability, 
causality and the second principle of thermodynamics. We analyze in detail time reversibility of a paradigmatic dissipative nonlinear dynamical system, namely the logistic map $x_{t+1}=1-ax_t^2$. 
A close relation is revealed between time reversibility and the sensitivity to the initial conditions. Indeed, depending on the initial condition and the size of the time series, time reversal can enable the recovery, within a small error bar, of past information when the Lyapunov exponent is non-positive, notably at the Feigenbaum point (edge of chaos), where weak chaos is known to exist. Past information is gradually lost for increasingly large Lyapunov exponent (strong chaos), notably at $a=2$ where it attains a large value. These facts open the door to diverse novel applications in physicochemical, astronomical, medical, financial, and other time series.  
\end{abstract}
%
\maketitle
The arrow of time has crucial implications in relevant issues related to causality, validity of the second principle of thermodynamics, reconstruction of a specific trajectory of a system, among others (see \cite{Zadunaisky1979} and references therein).
For example,
conservative Newtonian systems are deterministic and reversible with time. Indeed, the equation of motion $\vec{F}(\vec x)
=m\frac{d^2 \vec x}{dt^2}$ with given arbitrary initial conditions $(\vec x(0), \dot {\vec x}(0))$ yields an unique solution which is invariant through time reversal $t \to -t$. More precisely, assuming that all calculations are performed with infinite precision, we may reverse time at any moment including of course the sign change of all velocities $\dot {\vec x}$, and we obtain backwards the same trajectory that we had followed forward.

A different behavior is observed for paradigmatic nonlinear dynamical systems such as the dissipative logistic map 
\begin{equation}
x_{t+1}=1-ax_t^2\;\; (a \in [0,2]; \, x_t \in [-1,1];\,t=0,1,2,\dots)\,.
\label{logistic}
\end{equation}
Similarly to the previous Newtonian system, this map is deterministic in the sense that, once again assuming that we perform all calculations with infinite precision, the trajectory within the interval $[1,-1]$ is unique given an arbitrary initial condition $x_0$. However, Eq. (\ref{logistic}) is not time-reversible. Indeed, for any given $x_{t+1}$ we obtain two, and not one, values for $x_t$.

For both classes of systems, the concept of sensitivity to the initial conditions is, nevertheless, applicable. Indeed, the time evolution of a pair of different, infinitely close, initial conditions might exhibit (i) exponential divergence with time (i.e., positive Lyapunov exponents), thus causing strong chaos; (ii) exponential convergence with time (i.e., negative Lyapunov exponents), thus characterizing stable orbits; (iii) subexponential evolution with time (i.e., zero Lyapunov exponents). The latter corresponds, in many occasions, to the edge of chaos, thus reflecting weak chaos. The corresponding subexponential divergence with time is present in many natural, artificial and social complex systems and, for vast classes of phenomena, it is of the power-law type.

In the present paper we reveal a --- to the best of our knowledge --- new connection between time reversibility and sensitivity to the initial conditions. We focus on the logistic map as an illustrative case. Let us start with two paradigmatic values of the control parameter $a$, namely $a=2$ (occasionally referred to as the {\it Ulam point}), whose Lyapunov exponent $\lambda$ is given by $\lambda=\ln 2 >0$, thus corresponding to strong chaos, and $a=a_c\equiv 1.40115518909205\dots$ ({\it Feigenbaum-Coullet-Tresser point}), whose Lyapunov exponent $\lambda$ vanishes, corresponding to weak chaos.  

For the strong chaos case, the Kolmogorov-Sinai-like entropy production rate is characterized by the entropic functional introduced in the
pioneering works of Boltzmann and Gibbs \cite{Boltzmann1872,Boltzmann1877,Gibbs1901} 
\begin{equation}
S_{BG}=-k\sum_{i=1}^W p_i \ln p_i \;\; \Bigl(\sum_{i=1}^W p_i=1\Bigr)\,,
\label{BGentropy}
\end{equation} 
$k$ being a conventional positive constant adopted once forever (in physics, $k$ is chosen to be the Boltzmann constant $k_B$; in information theory and computational sciences, it is frequently adopted $k=1$ and logarithm to the base 2). Here we work with discrete random variables, but the arguments can straightforwardly be translated for continuous cases. 

In the simple case of equal probabilities, this entropic functional is given by $S_{BG}=k \ln W$. 
Moreover, Eq. ({\protect~\ref{BGentropy}}) is generically {\it additive} \cite{Penrose1970}. Indeed, if $A$ and $B$ are two probabilistically independent systems (i.e., $p_{ij}^{A+B}= p_i^A p_j^B$), we straightforwardly verify that
\begin{equation} 
S_{BG}(A+B)=S_{BG}(A)+S_{BG}(B) \,.
\end{equation} 
This celebrated entropic functional is consistent with thermodynamics for all systems whose $N$ elements are either independent or weakly interacting in the sense that only basically local (in space/time) correlations are involved. For example, if we have equal probabilities (typical microcanonical ensemble) and the system is such that the number of accessible microscopic configurations is given by $W(N) \propto \mu^N\; (\mu>1; \,N\to\infty)$, then $S_{BG}(N)$ is {\it extensive} as required by thermodynamics. 
Indeed $S_{BG}(N)=k\ln W(N) \sim k(\ln \mu)N$. But if the correlations are nonlocal in space/time, $S_{BG}$ may become thermodynamically inadmissible. Such is the case of equal probabilities with say $W(N) \propto N^\nu \;(\nu>0; \,N\to\infty)$: it immediately follows $S_{BG}(N) \propto \ln N$, which violates thermodynamical extensivity. 
To satisfactorily approach cases such as this one, it was proposed in 1988 \cite{Tsallis1988,TsallisMendesPlastino1998,GellMannTsallis2004,Tsallis2009} to build a more general statistical mechanics based on the {\it nonadditive} entropic functional
\begin{equation}
S_q\equiv k\frac{1-\sum_{i=1}^W p_i^q}{q-1}=k\sum_{i=1}^W p_i \ln_q \frac{1}{p_i} = -k\sum_{i=1}^W p_i^q \ln_q p_i = -k\sum_{i=1}^W p_i \ln_{2-q} p_i \;\;(q \in \mathbb{R}; S_1=S_{BG})\,,
\end{equation}
with $\ln_q z \equiv \frac{z^{1-q}-1}{1-q} \; (\ln_1 z=\ln z)$ and its inverse $e_q^z \equiv [1+(1-q)z]_{+}^{1/(1-q)}$; $(e_1^z=e^z$; $[z]_{+}=z$ if $z>0$ and vanishes otherwise); for $q<0$, it is necessary to exclude from the sum the terms with vanishing $p_i$. We easily verify that equal probabilities yield $S_q=k\ln_q W$, and that generically we have
\begin{equation}
S_q(A+B)=S_q(A)+S_q(B)+\frac{1-q}{k}S_q(A)S_q(B) \,.
\end{equation}
Consequently, in the $(1-q)/k \to 0$ limit, we recover the $S_{BG}$ additivity. For the anomalous class of systems mentioned above, namely $W(N) \propto N^\nu$, we obtain, $\forall \nu$,  the {\it extensive} entropy $S_{1-1/\nu}(N)=k\ln_{1-1/\nu}W(N) \propto N$, as required by the Legendre structure of thermodynamics (see \cite{Tsallis2009,TsallisCirto2013,Tsallis2022} and references therein). 

Let us remind at this stage the connection between the entropy production rate $K_{BG}$ and the Lyapunov exponent $\lambda_{1}$. For the $a=2$ logistic map, the following Pesin-like identity holds \cite{LatoraBaranger1999}
\begin{equation}
K_{BG} \equiv \lim_{t\to\infty} \lim_{W\to\infty} \lim_{M\to\infty} \frac{\langle S_{BG}\rangle(t)}{t}= \langle \lambda_1\rangle \,,
\label{Pesin}
\end{equation}
where we have partitioned the phase space $x\in [-1,1]$ in $W \gg 1$ little cells, and have started the map evolution by randomly putting $M\gg 1$ (typically $M=10 W$) initial conditions within one of those cells, the averaging being done over all $W$ cells; the Lyapunov exponent is defined through the sensitivity to the initial conditions $\xi(t) \equiv \frac{\Delta x(t)}{\Delta x(0)}=e^{\lambda_1\,t}$ \,.

Eq. (\ref{Pesin}) can be generalized as follows
\begin{equation}
K_{q} \equiv \lim_{t\to\infty} \lim_{W\to\infty} \lim_{M\to\infty} \frac{\langle S_{q}\rangle(t)}{t}= \langle \lambda_q\rangle \,,
\label{qPesin}
\end{equation}
where $\lambda_q$ is defined through $\xi(t)=e_q^{\lambda_q\,t} [\sim t^{1/(1-q)}$ if $q<1$] for $t\gg 1$ and $q\le 1$. For the $a=a_c$ logistic map, the following $q$-generalized Pesin-like identity holds \cite{TsallisPlastinoZheng1997,LyraTsallis1998,SilvaCruzLyra1999,LatoraBarangerRapisardaTsallis2000,BaldovinRobledo2002,AnanosTsallis2004,BaldovinRobledo2004}:
\begin{equation}
K_{q_c} = \langle \lambda_{q_c}\rangle \,,
\label{qcPesin}
\end{equation}
with $q_c=0.2445\dots $ ($10,026$ meaningful digits are presently known).

The {\it finiteness} of $K_q$ imposes a drastic change in the value of $q$ when we contrast the $a=2$ case (for which we have a smooth $t\to\infty$ attractor and, consistently, $q=1$) with the $a=a_c$ case (for which we have a multifractal $t\to\infty$ attractor and, consistently, $q=q_c$). This fact raises a crucial question, namely {\it what happens with time reversibility?} In what follows we reveal a sensible difference between these two paradigmatic cases. 

\begin{figure}[h]
\begin{minipage}{0.47\linewidth}
 \includegraphics[scale=0.40,keepaspectratio,clip=]{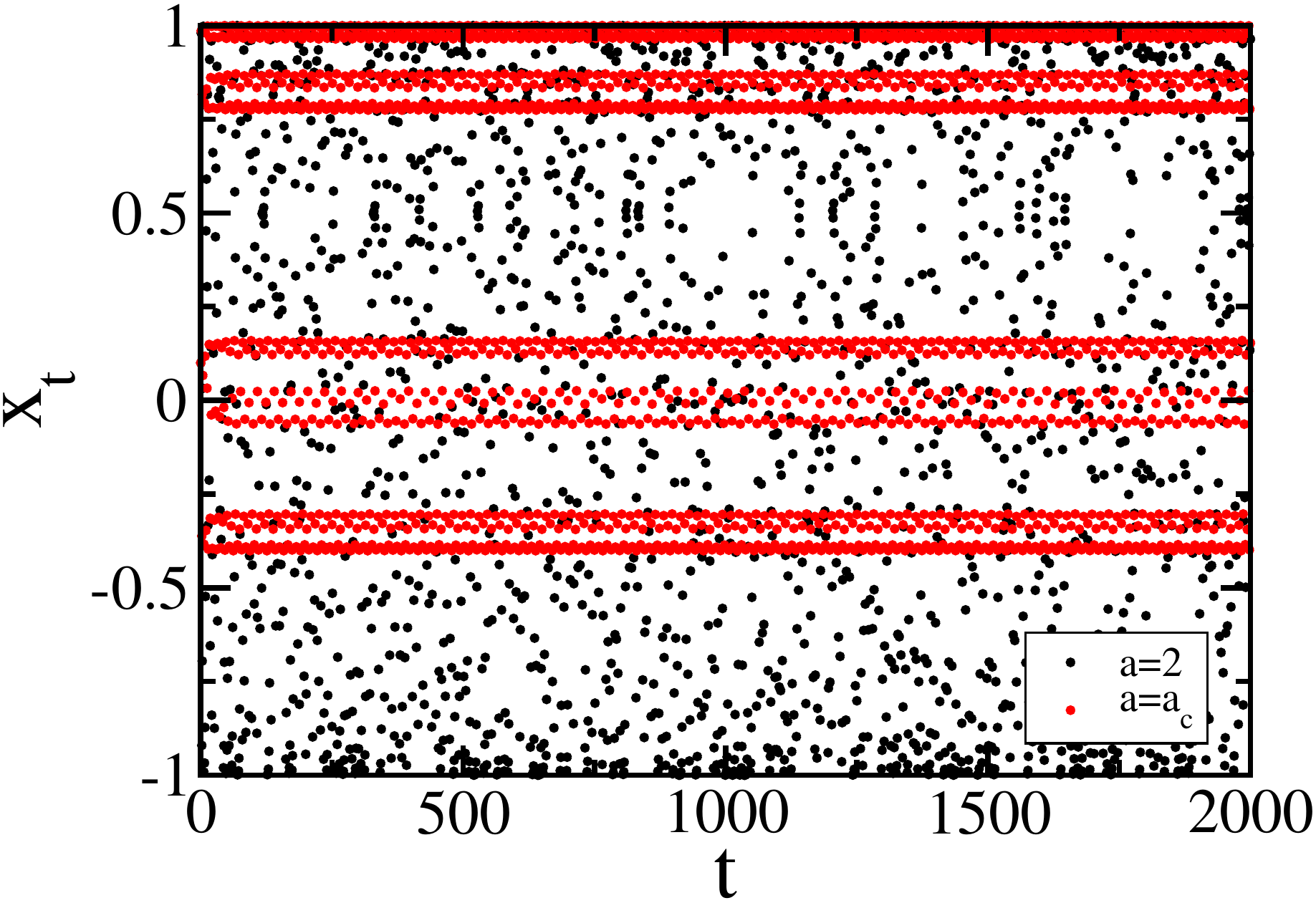}
\end{minipage}
\hspace{-0.6cm}
\begin{minipage}{0.47\linewidth}
 \hspace{0.7cm}
 \includegraphics[scale=0.40,keepaspectratio,clip=]{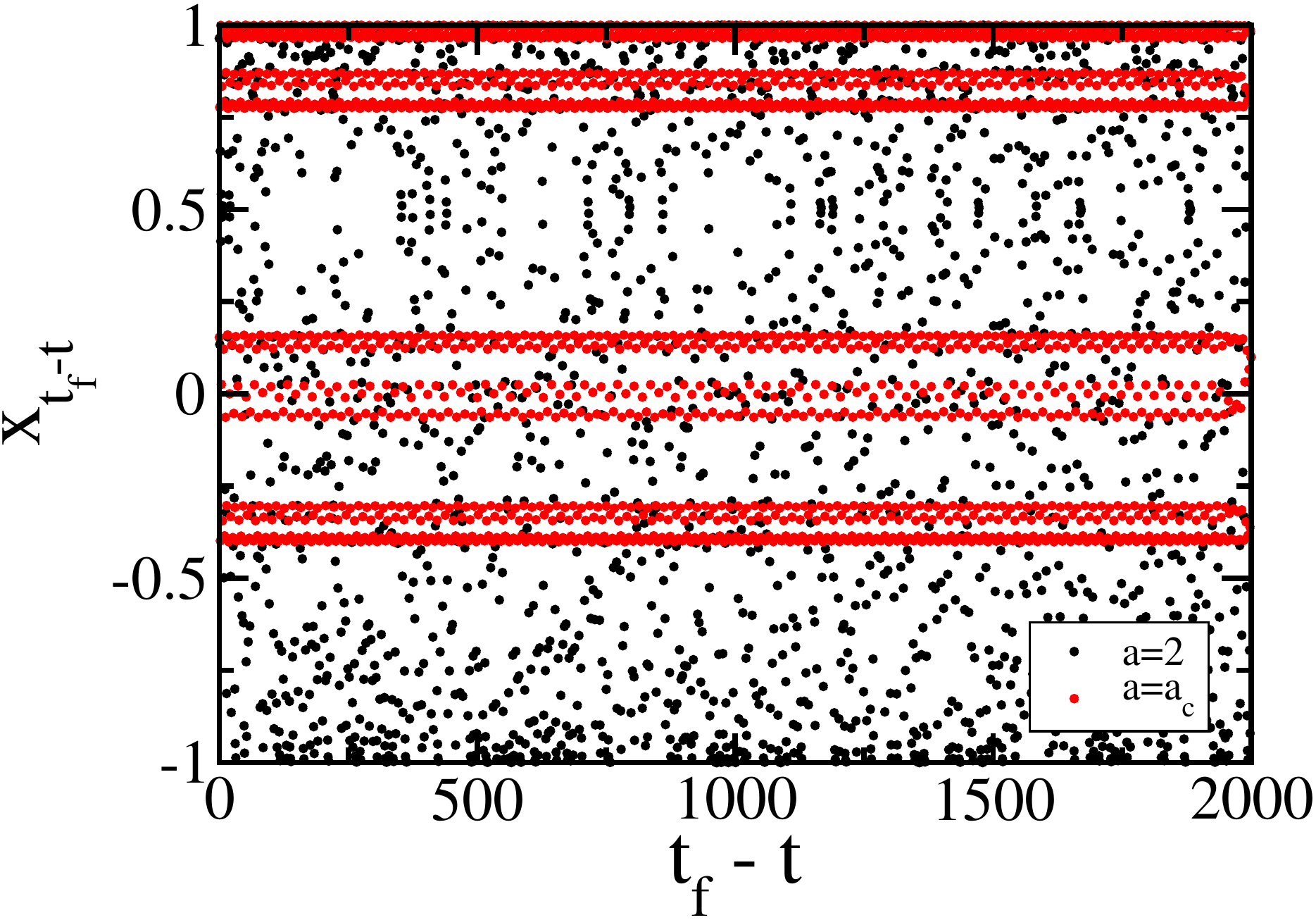}
\end{minipage}
\\
\begin{minipage}{0.47\linewidth}
 \includegraphics[scale=0.40,keepaspectratio,clip=]{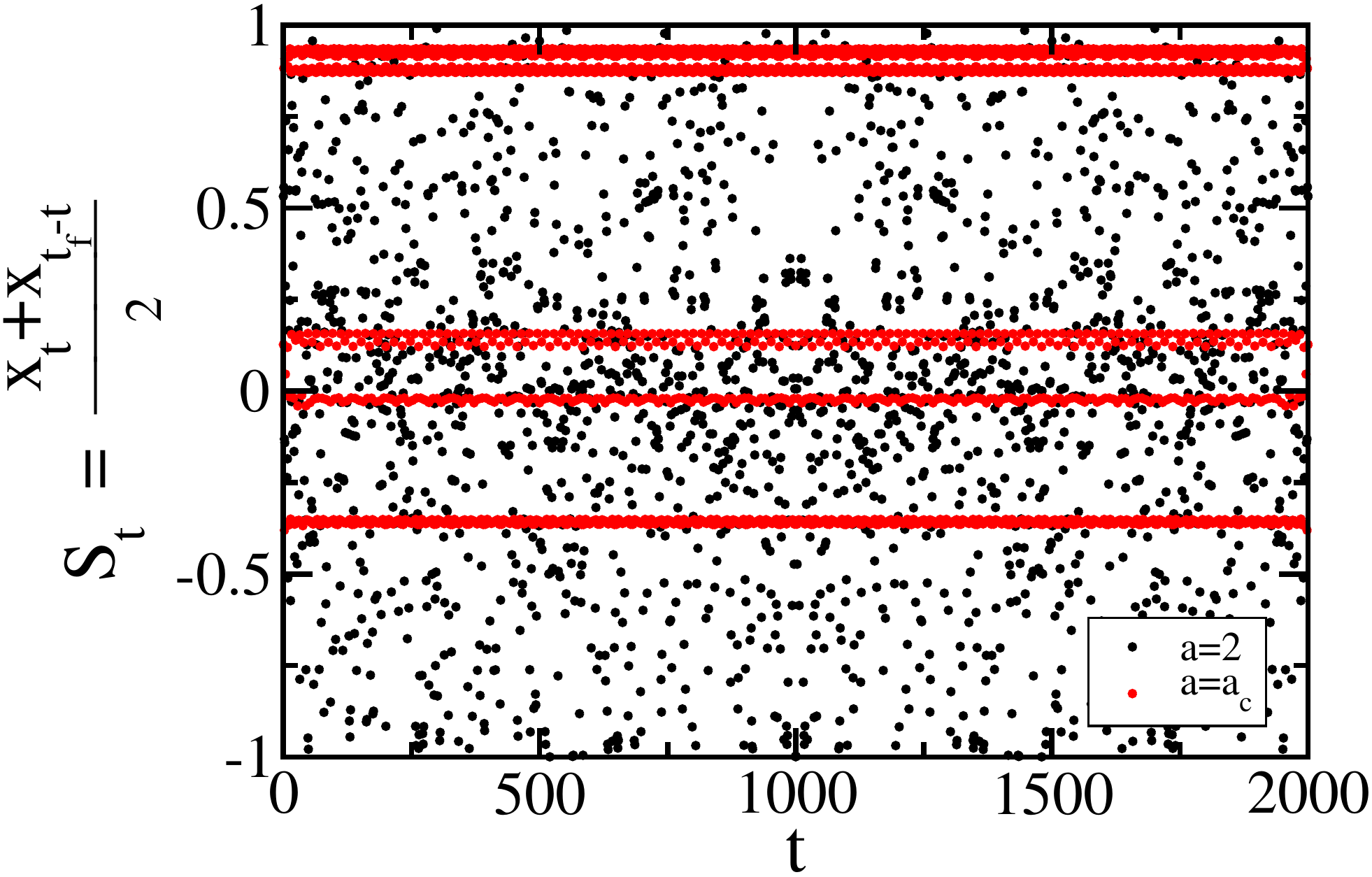}
\end{minipage}
\begin{minipage}{0.47\linewidth}
 \includegraphics[scale=0.40,keepaspectratio,clip=]{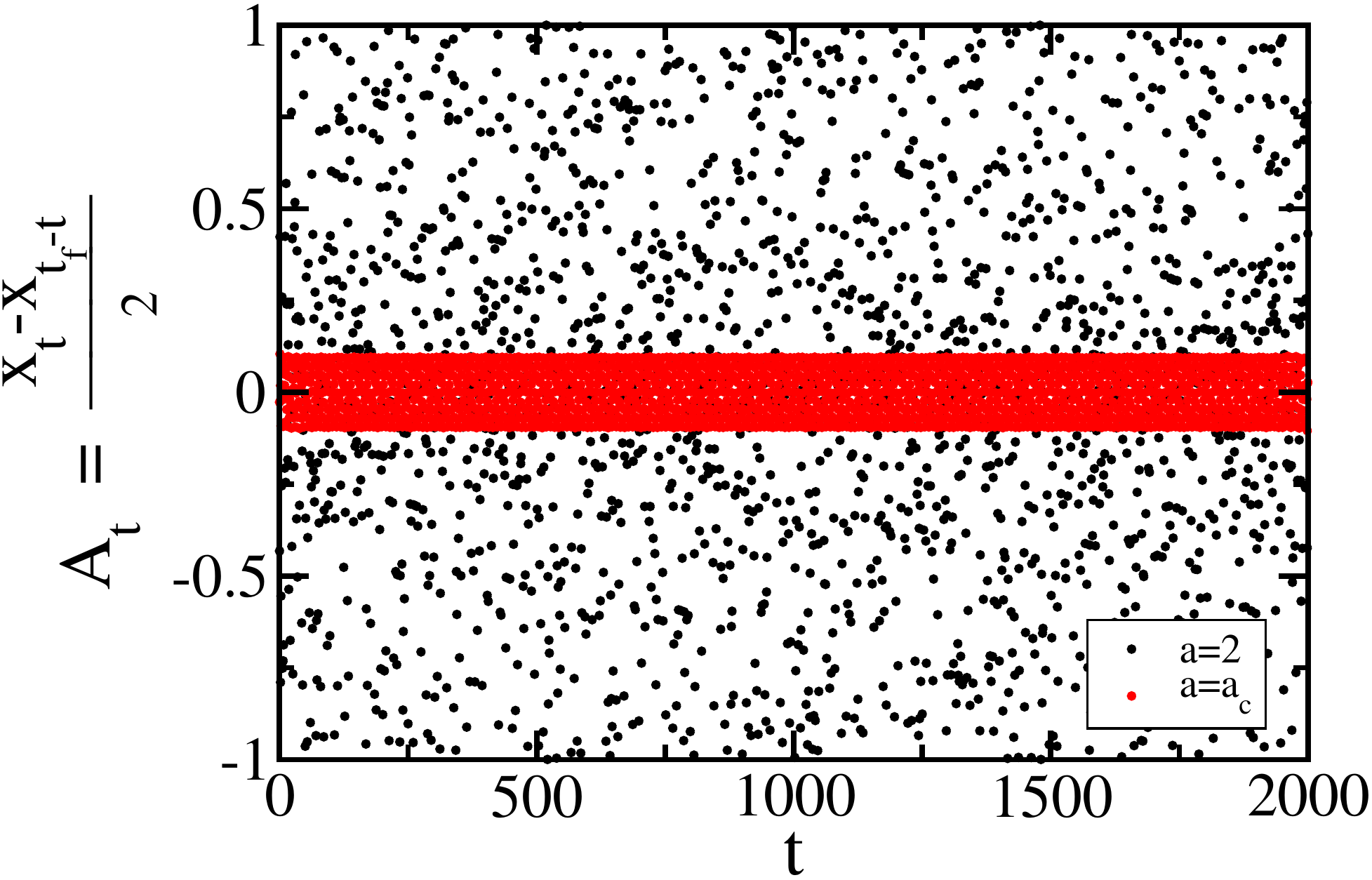}
\end{minipage}
\begin{minipage}{0.47\linewidth}
 \includegraphics[scale=0.40,keepaspectratio,clip=]{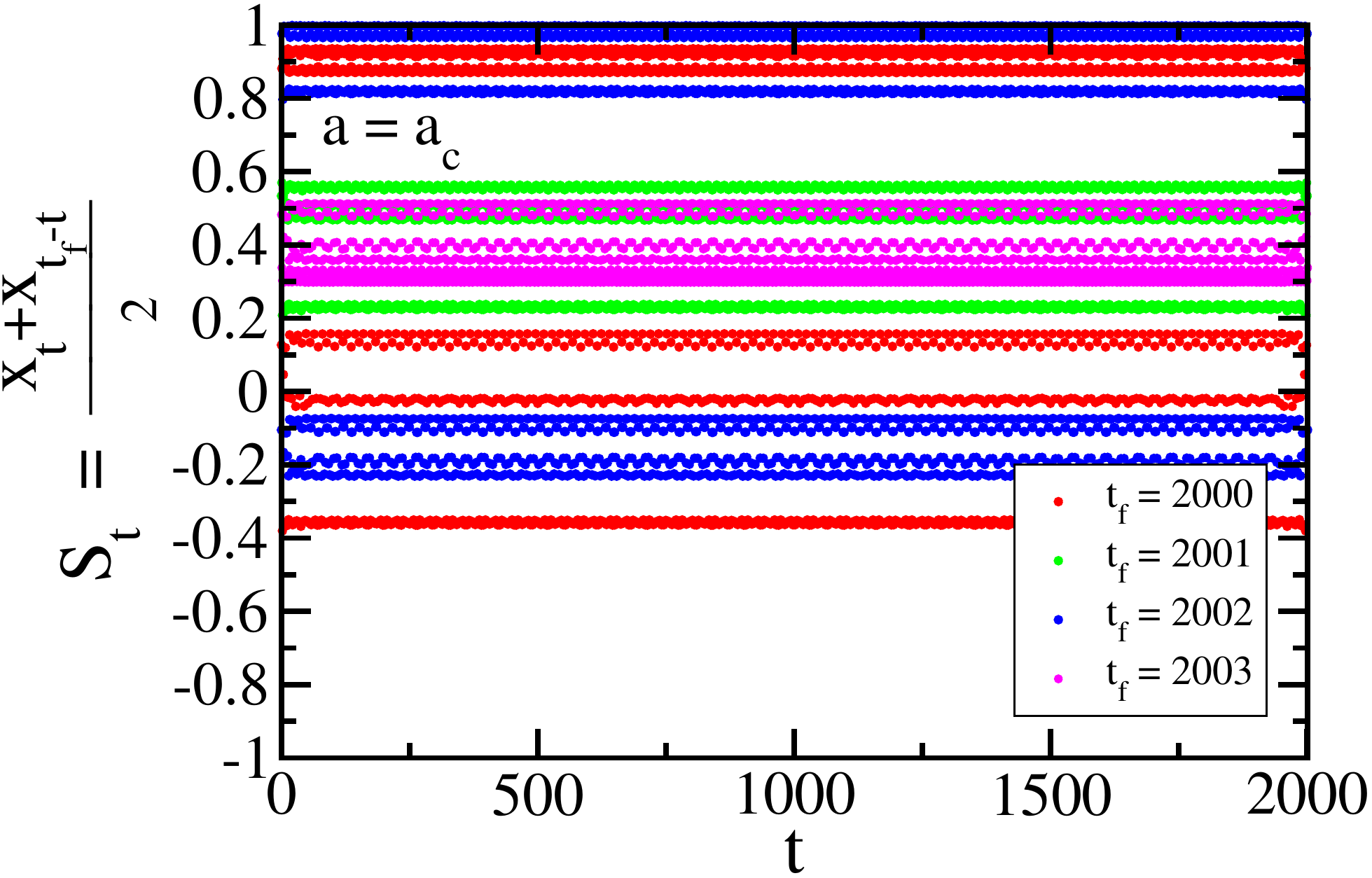}
\end{minipage}
\begin{minipage}{0.47\linewidth}
 \includegraphics[scale=0.40,keepaspectratio,clip=]{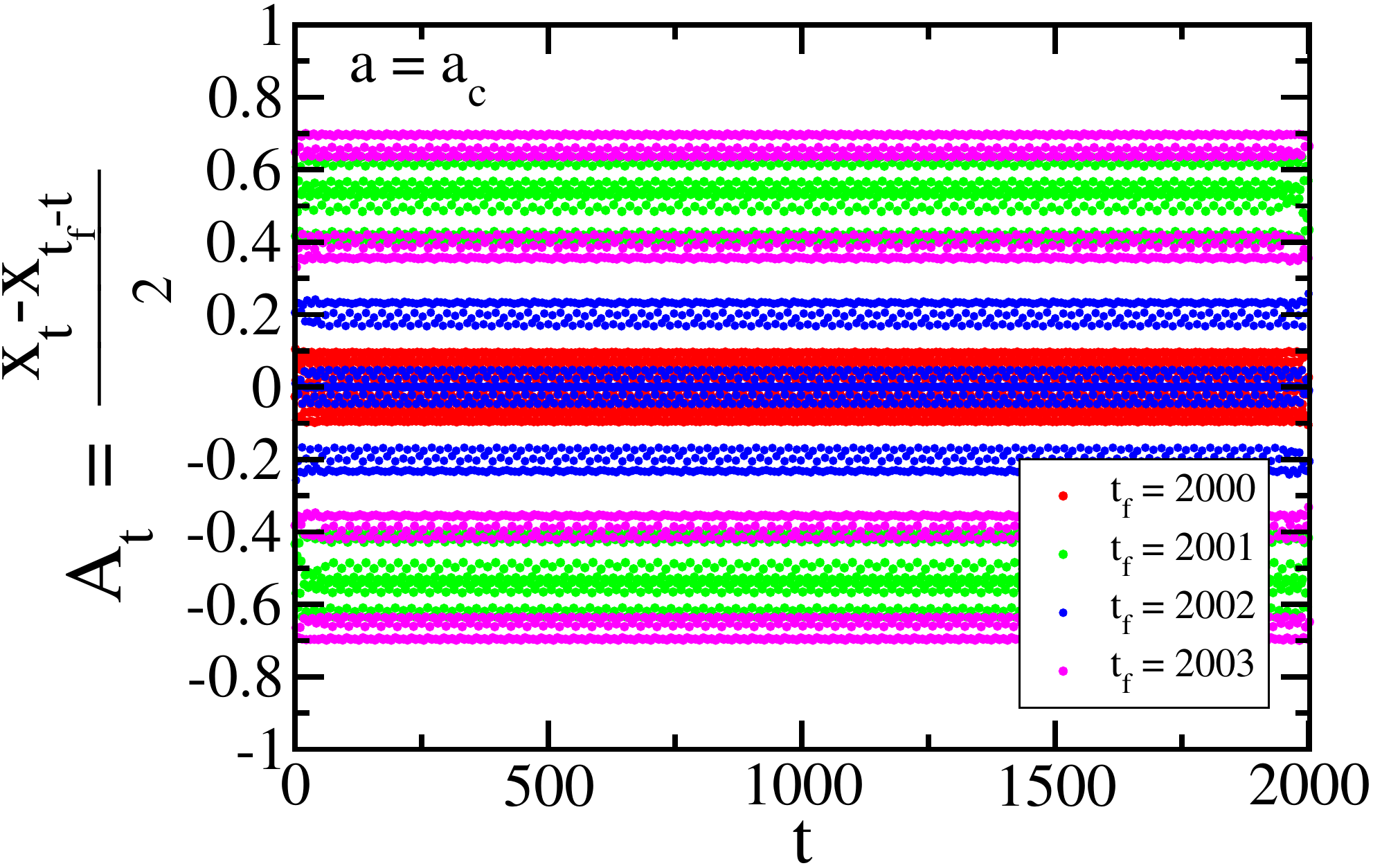}
\end{minipage}
\caption{\small (Color online) 
         {\it Top plots:} Time dependence of $x_t$ and of $x_{t_f-t}$ with $x_0=0.1$ and
         $t_f$=2000 for $a=2$ (black dots) 
         and $a=a_c \equiv 1.40115518909205\dots$ (red dots).
         {\it Middle plots:} Time dependence of $S_t$ and $A_t$.
         {\it Bottom plots:}  The value of $t_f$ has relevant consequences, as illustrated here.} 
\label{fig1}
\end{figure}

\begin{figure}[h]
\begin{minipage}{1.00\linewidth}
\hspace{-0.9cm}
 \includegraphics[scale=0.40,keepaspectratio,clip=]{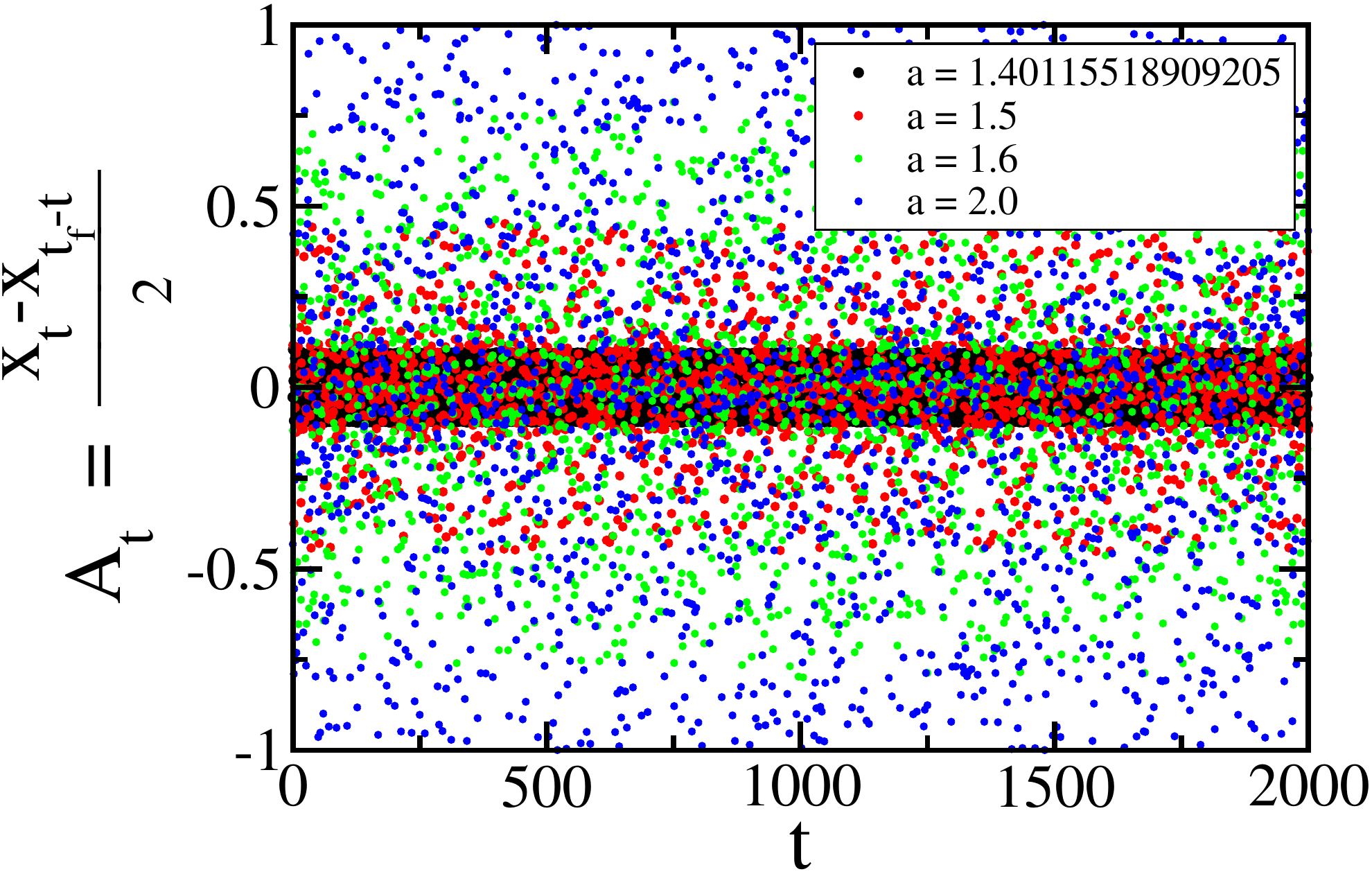}
\end{minipage}

\begin{minipage}{0.47\linewidth}
\hspace{0.75cm}
\includegraphics[scale=0.42,keepaspectratio,clip=]{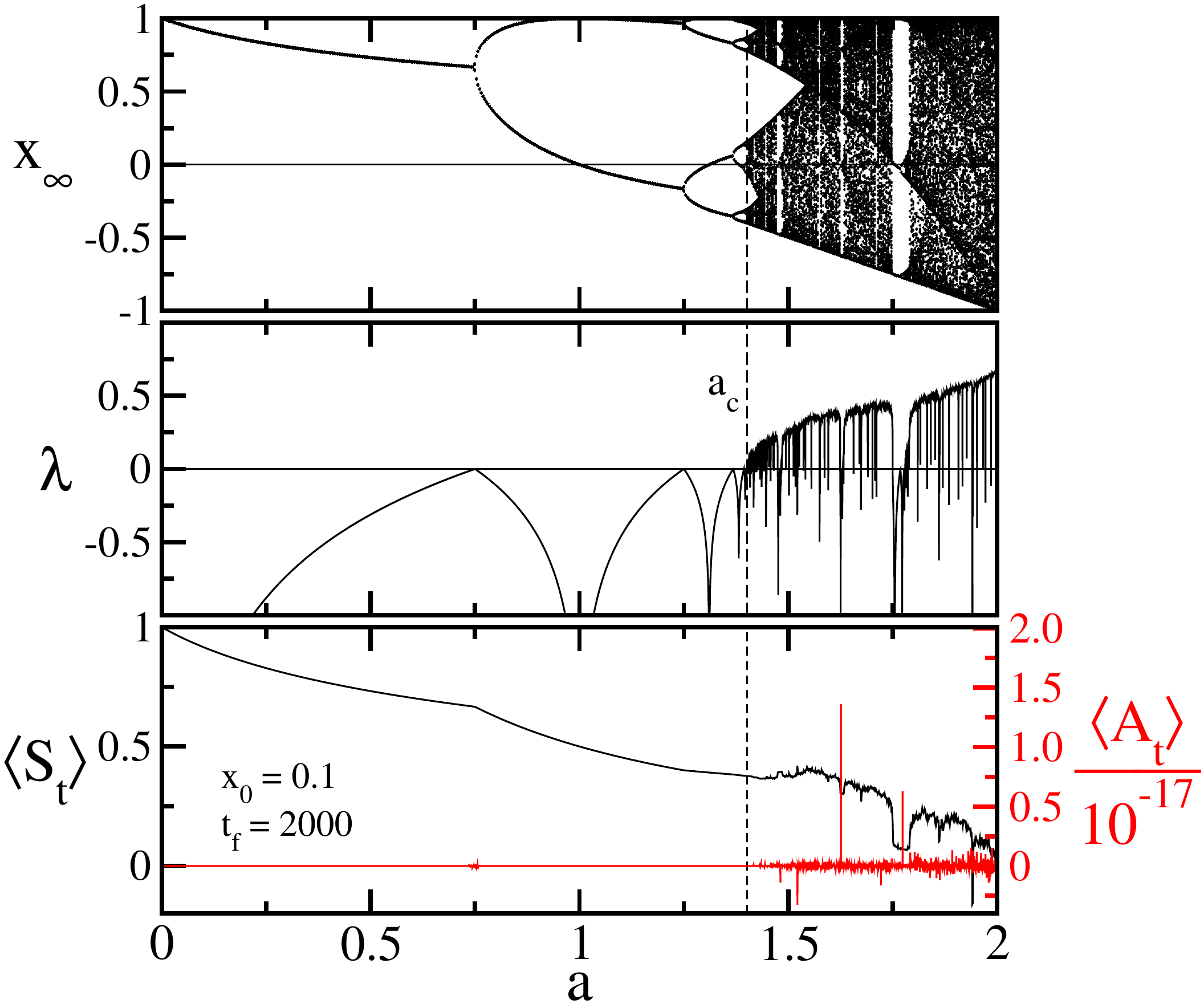}
\end{minipage}
\\
\caption{\small (Color online)  
         {\it Top:} Influence, for $(x_0,t_f)=(0.1,2000)$, of $a$ on $A_t$ [$A_t$ gradually (though non monotonically) 
         spreads within the entire phase space  interval $[1,-1]$ 
         for $a$ increasing from $a=a_c$ to $a=2$]; 
         {\it Bottom:} The influence of $a$ on $\langle S_t\rangle$ and $\langle A_t\rangle$ for $(x_0,t_f)=(0.1,2000)$; for other values of $(x_0,t_f)$, $\langle S_t \rangle$ and $\langle A_t \rangle$ remain nearly the same.
}
\label{fig2}
\end{figure}

We start with an arbitrarily chosen $x_0$ and iterate the map until arrival to a final value $t_{f}$. We then consider
\begin{equation}
x_t=\frac{x_t +x_{t_f-t}}{2}+ \frac{x_t -x_{t_f-t}}{2} \equiv S_t+A_t\,,
\end{equation}
where $S_t$ ($A_t$) is symmetric (antisymmetric) with regard to $t \leftrightarrow (t_f-t)$.  These quantities are depicted in Fig. \ref{fig1}, and we easily appreciate the considerable difference of $A_t$ for $a=2$ and $a=a_c$. Indeed, for $a=2$, the values of $A_t$ spread over the entire phase space whereas, for $a=a_c$, $A_t$ can be strongly concentrated around $A_t=0$ ! In other words, quasi-reversibility may occur for a multifractal attractor, whereas by no means such remarkable property is observed when the attractor includes smooth regions. The influence of $a$ on various relevant quantities is depicted in Fig. \ref{fig2}. 

\begin{figure}[h]
\centering
 \includegraphics[scale=0.40,keepaspectratio,clip=]{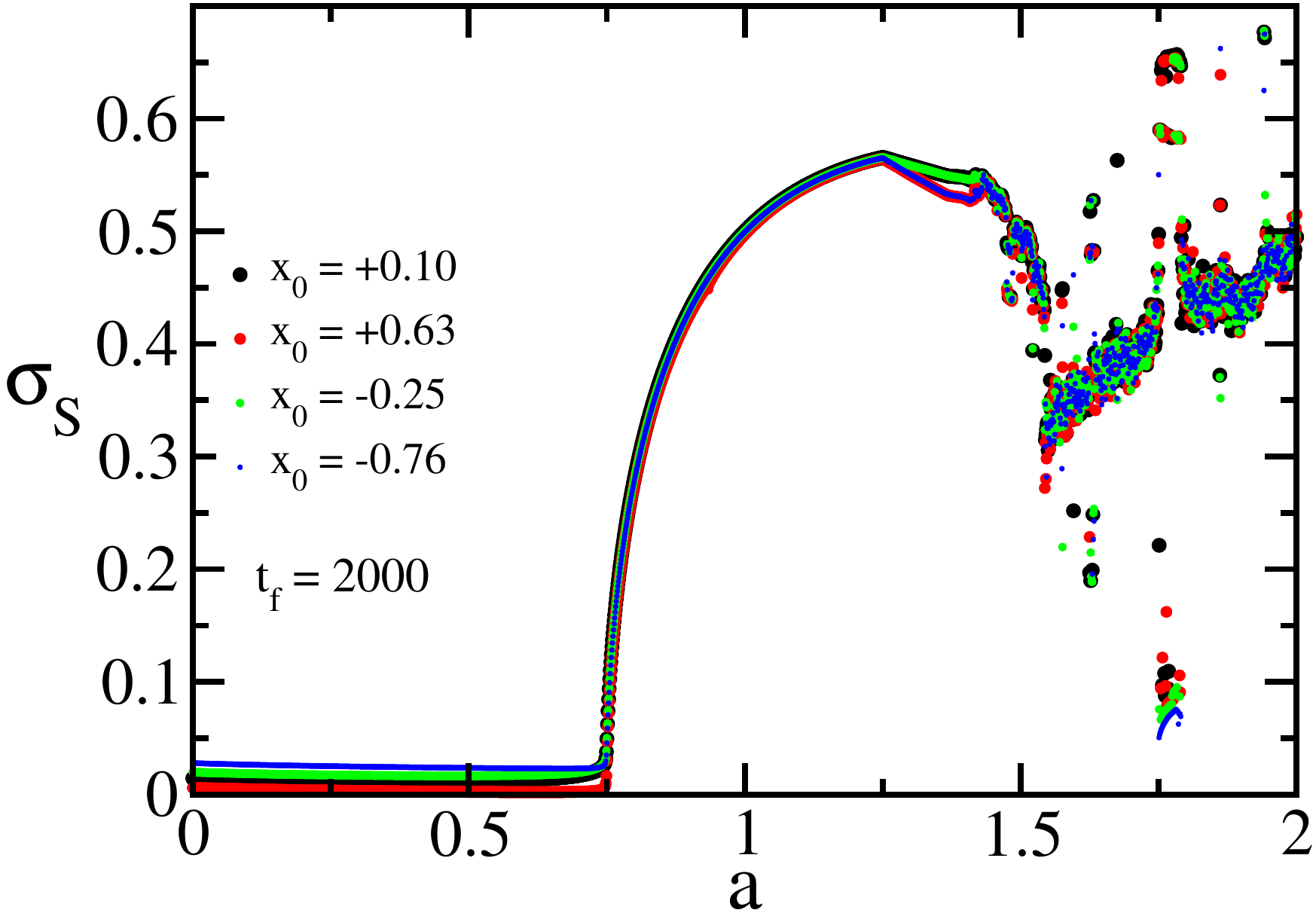}
 \includegraphics[scale=0.40,keepaspectratio,clip=]{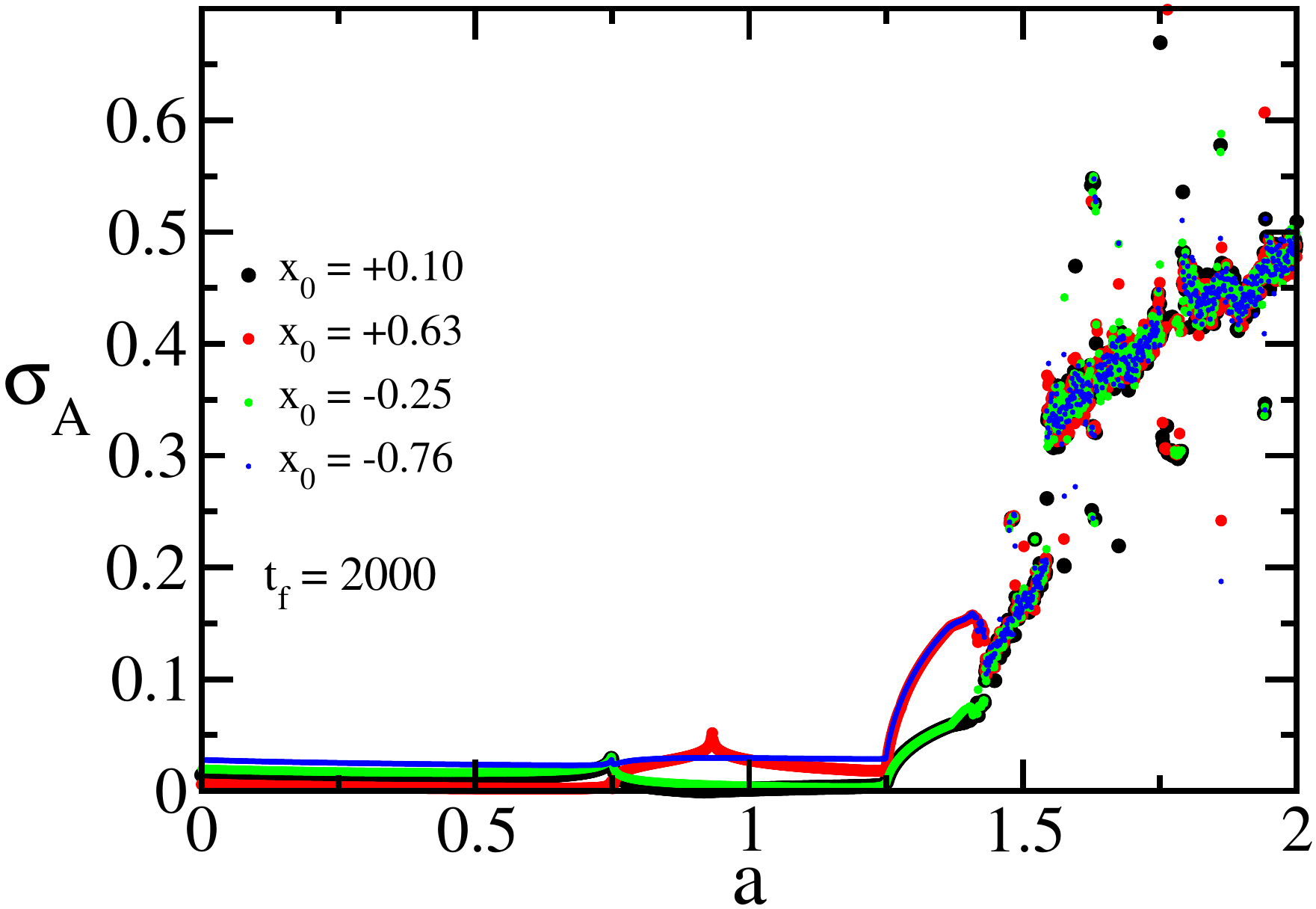}
\caption{\small (Color online) 
Influence of the initial condition on the standard deviations $\sigma_S$ and $\sigma_A$ for fixed $t_f$.
}
\label{figinitcond} 
\end{figure}
\begin{figure}[h]
\centering
  \includegraphics[scale=0.40,keepaspectratio,clip=]{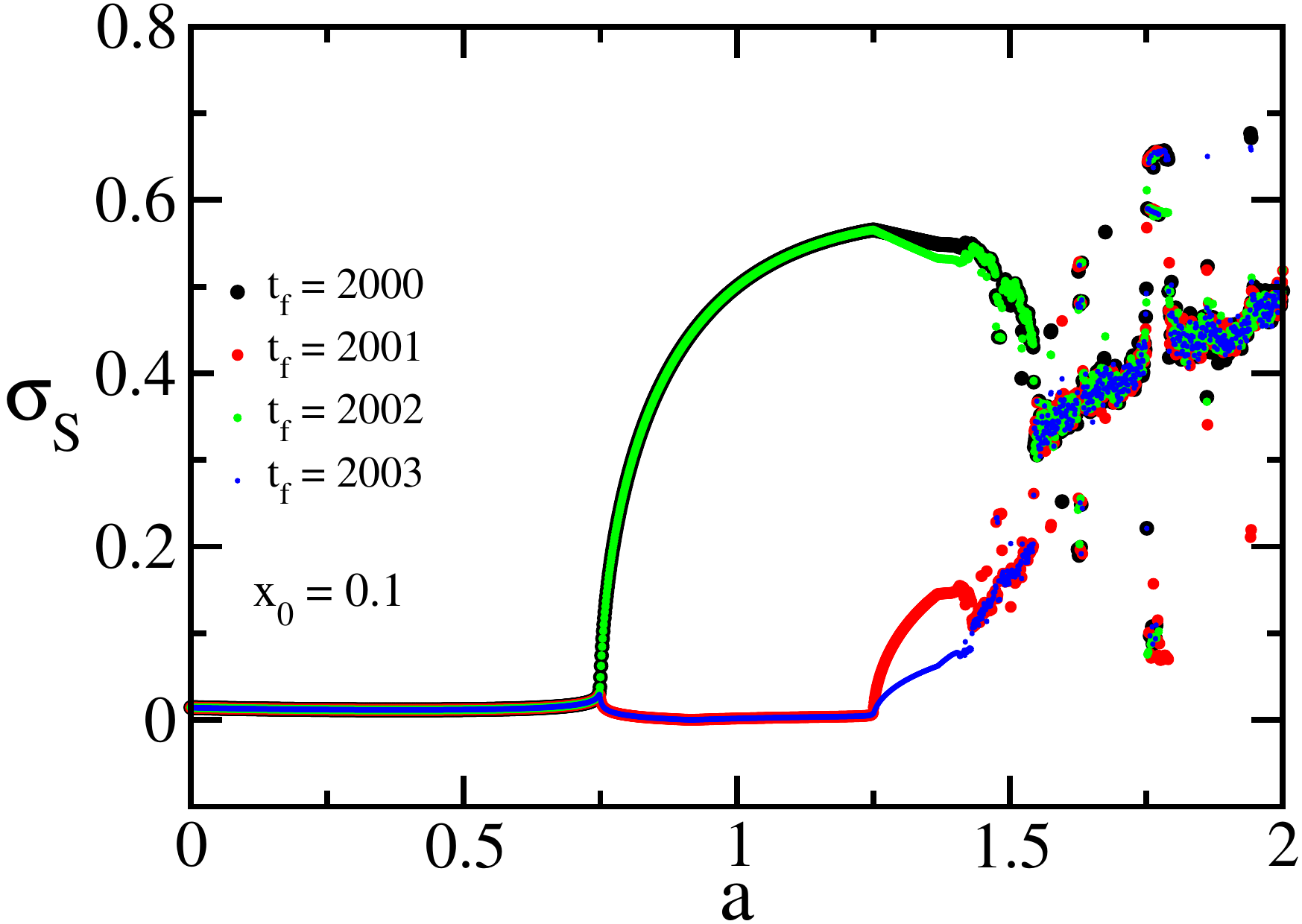}\\
  \includegraphics[scale=0.40,keepaspectratio,clip=]{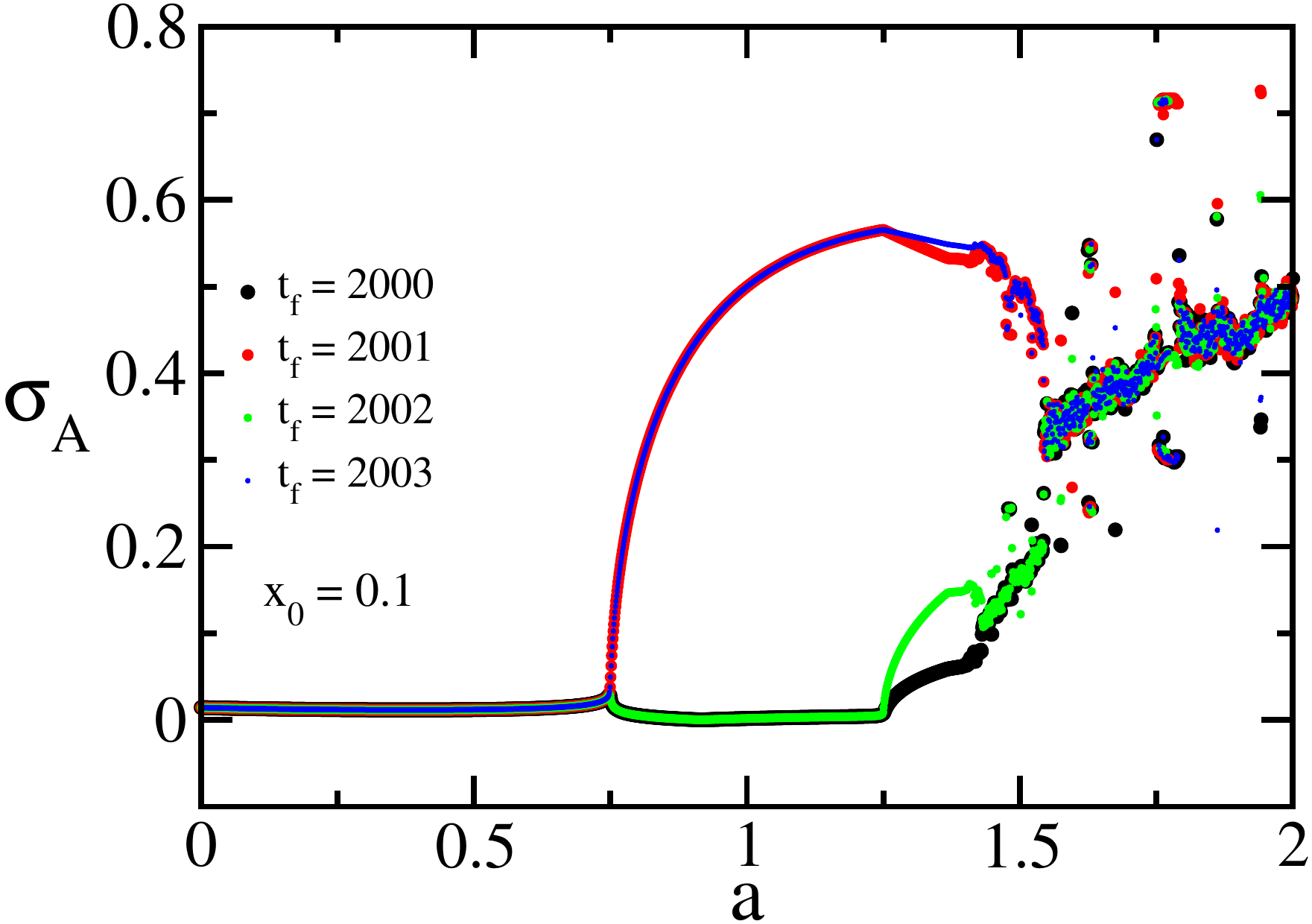}\\
 \includegraphics[scale=0.40,keepaspectratio,clip=]{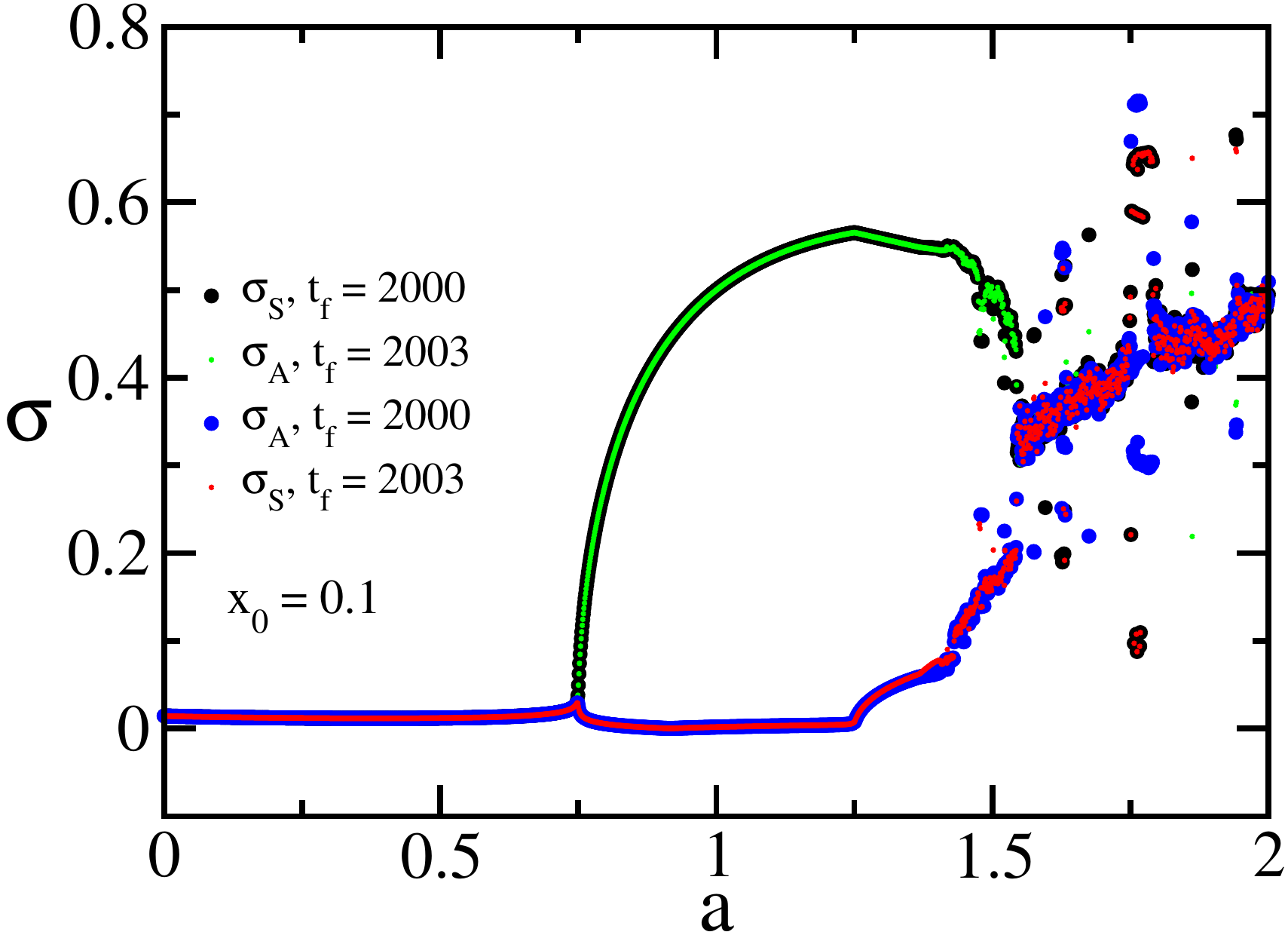}  
\caption{\small (Color online) {\it Top and Middle:} Influence of $(t_f,x_0)$ on $(\sigma_S,\sigma_A)$; for other values of $x_0$ the data follow similar paths.
{\it Bottom:} Illustration of the fact that, for exchanged values of $t_f$, $\sigma_S$ and $\sigma_A$ can nearly coincide.
}
\label{exchange} 
\end{figure}

Typical results of the influence of $(x_0,t_f)$ on the standard deviations $(\sigma_S,\sigma_A)$ are depicted in Figs. \ref{figinitcond} and \ref{exchange}.

Let us conclude by reminding that classical mechanics is deterministic for both time increasing forward or increasing backwards (see \cite{Zadunaisky1979} and references therein). In contrast, other paradigmatic nonlinear dynamical systems such as the dissipative logistic map are deterministic only for time increasing forward, not so for time increasing backwards. We have here focused on this map because of its relevance for many phenomena in natural sciences, and because it is able to exhibit, in addition to stable orbits, both strong (positive Lyapunov exponent) and weak chaos (vanishing Lyapunov exponent), depending on the value of its external control parameter. These two classes of chaos illustrate  basic features of, respectively, Boltzmann-Gibbs statistical mechanics (grounded on the additive entropic functional $S_{BG}$) and nonextensive statistical mechanics (grounded on the nonadditive entropic functional $S_q$). Consistently, the entropy production rate which appears in the Pesin-like identity concerns $q=1$ for strong chaos (whose dynamical attractor includes regions of smooth functions) and $q=q_c=0.2445\dots$ at the Feigenbaum point (whose dynamical attractor is known to be a complex multifractal). On the basis of these quite different dynamical behaviors, we revealed here that quasi-reversibility with time, i.e. $x_{t_f-t} \simeq x_t$, is possible when the Lyapunov exponent is nonpositive (at the Feigenbaum point, for instance), whereas it is essentially lost when the Lyapunov exponent is positive. This connection between a global time behavior, namely reversibility, and the sensitivity to the initial conditions might be very useful for analyzing issues such as predictability, causality and others in physicochemical, astrophysical, economical and medical time series (e.g., EEG and ECG), among others. In fact, the practical usefulness of time-reversal has been very recently exhibited by decreasing error bars in the predictions of strong earthquakes \cite{Varotsosetal2023}.
The study of how general this phenomenon might be, and for what specific classes of low-dimensional and/or high-dimensional nonlinear dynamical systems, certainly constitutes a deeply interesting challenge, to be addressed in future efforts.

We acknowledge fruitful remarks by D.M. Abramov, E.M.F. Curado, H.S. Lima, A.R. Plastino and R. Wedemann, as well as partial financial support by CNPq and Faperj (Brazilian agencies).

\end{document}